\documentstyle[11pt]{article}
\begin{document}

\title{Phase
Information in Quantum Oracle Computing} 

\author{Jonathan Machta
\thanks{This research was partially funded by the National
Science Foundation Grant DMR-9311580.}\\
Department of Physics and Astronomy\\
University of Massachusetts\\
Amherst, Massachusetts 01003\\
e-mail address: machta@phast.umass.edu
}
\date{May 31, 1994}
\maketitle

\begin{abstract}
Computational devices may be supplied with external sources of
information (oracles). Quantum oracles may transmit phase
information which is available to a quantum computer but not a
classical computer. One consequence of this observation is that
there is an oracle which is of no assistance to a classical computer
but which allows a quantum computer to solve undecidable problems.
Thus useful relativized separations between quantum and classical
complexity classes must exclude the transmission of phase
information from oracle to computer.   

\end{abstract}

Deutsch~\cite{De85} introduced the notion of a universal quantum
computer. This machine is the natural generalization of a
(reversible) Turing machine to the quantum world.  The universal
quantum computer differs from its classical counterpart in that it
may evolve in a superposition of states.  Thus there is the prospect
that a single quantum computer may be able to carry out many
simultaneous calculations.  Unfortunately, the naive idea of quantum
parallelism is defeated by the need to make a measurement at the end
of the computation so that only one of the many parallel
computations is available to the (classical) user. 
Nonetheless, a more restricted notion of quantum parallelism was
recently introduced by Deutsch and Jozsa~\cite{DeJo}.  They show
that relative to an `oracle', a quantum computer may solve certain
problems much faster than a Turing machine equipped with the same
oracle.  An oracle is an auxiliary device to which the computer can
address queries and receive YES or NO answers.  Deutsch and Jozsa
envisioned a quantum oracle which may receive a superposition of
questions and return a superposition of answers.  Later Berthiaume
and Brassard \cite{BeBr92,BeBr93} refined the notion of `oracle
quantum computing' and obtained results separating  conventional
(classical) and quantum complexity classes relative to appropriate
oracles.

It is the purpose of this note to investigate the role of phase
information in the interaction between a quantum computer and an
oracle.  The answers provided by a quantum oracle contain both
amplitude and phase information.   In \cite{DeJo} it is assumed that
the oracle increments the phase of the wave function by the same
amount for all queries.  However, there is no physical reason for
this choice.  If we allow for full use of phase information a
quantum oracle can transmit information to a quantum computer which
is inaccessible to a classical computer equipped with the same
oracle.  
 
The conventional computer science definition of an oracle is a set
$X$, $X \subseteq \Sigma^\ast$ where $\Sigma^\ast$ is the set
of all finite bit strings.  A classical realization of an oracle is a
device which when fed a string $x \in \Sigma^\ast$ (the query)
returns a `$1$' if $x\in X$ and `$0$' otherwise.  

The states of a quantum mechanical system are described as vectors in
a Hilbert space and the evolution of the system by unitary
transformations on the states.  To simplify the discussion suppose
that the quantum computer and oracle device act on a
finite dimensional Hilbert space and that the abstract oracle takes
the form  $X \subseteq \Sigma^n$ where $\Sigma^n$ is the set of all
$n$-bit strings.  In Dirac notation, a basis for the Hilbert space
is given by $|x,y \rangle$ where $x \in \Sigma^n$ and $y$ is a
single bit which represents the answer to the query $x$.
A quantum oracle is naturally viewed as a unitary transformation, $U$
of the form  
\begin{equation}
| x,0 \rangle \stackrel{U}{\rightarrow}
e^{i\phi_{x,0}}| x,f(x) \rangle
\end{equation}
 where $f(x)=1$ if $x \in X$ and
$f(x)=0$ otherwise. Note
that unitarity implies that 
\begin{equation}
| x,1 \rangle
\stackrel{U}{\rightarrow} e^{i\phi_{x,1}}| x,1+f(x) \;{\rm mod} 2
\rangle
\end{equation}
This definition of $U$ differs from the one given in Eq. (2) of
\cite{DeJo} because of the arbitrary phase factor
$e^{i\phi_{x,y}}$.  The more restrictive definition
introduced in \cite{DeJo} assumes that the all the phase factors are
the same.  It must be stressed that there is no physical reason for
making this choice. In general, a quantum device which faithfully 
represents the abstract oracle, $X$ may be defined with any choice
for the phase angles $\phi_{x,y}$ and, unless special care is taken,
there is no reason to expect that the phase angles will all be
equal.

To understand the role of phase, we briefly review the  computation
described in \cite{DeJo}.  The quantum computer is prepared in a
superposition of the form 
\begin{equation} | \psi \rangle =
\frac{1}{\cal N} \sum_x | x,0 \rangle \end{equation} where ${\cal
N}$ is an appropriate normalization.  This state is fed to the
oracle which returns the state 
\begin{equation} U| \psi \rangle =
\frac{1}{\cal N} \sum_x  e^{i\phi_{x,0}} | x,f(x) \rangle
\end{equation} 
The operation of the computer effects the
transformation, $S$,  
\begin{equation}
| x,y \rangle \stackrel{S}{\rightarrow} (-1)^y|
x,y \rangle
\end{equation}
After the computer runs the state is given by
\begin{equation}
SU| \psi \rangle = \frac{1}{\cal N} \sum_x 
(-1)^{f(x)}e^{i\phi_{x,0}} | x,f(x) \rangle
\end{equation}
Finally, this state is again transformed by the oracle yielding
\begin{equation}
| \chi \rangle =USU| \psi \rangle = \frac{1}{\cal N} \sum_x 
(-1)^{f(x)}e^{i(\phi_{x,0}+\phi_{x,f(x)})} | x,0 \rangle
\end{equation}
Suppose that all of the phase angles are zero and that one
of the following two properties hold for the oracle: (A) All
strings are accepted by the oracle ($X=\Sigma^n$) or (B) Exactly half
of all strings are accepted by the oracle ($|X|=2^{n-1}$).  Given
these assumptions, Deutsch and Jozsa show how to utilize quantum
parallelism to determine with certainty which of (A) or (B) holds.  
The inner product $\langle \psi | \chi \rangle$  is $1$ if (A) holds
whereas $\langle \psi | \chi \rangle =0$ if (B) holds.  A measurement
of the operator $| \psi \rangle \langle \psi |$ thus distinguishes
between these two possibilities. It is shown in \cite{DeJo} this can
be accomplished in fewer computational steps than is possible given a
classical computer and the same oracle.  This method is the basis of
the relative separations between quantum and classical complexity
classes proved in \cite{BeBr92,BeBr93,BeVa}.  

There are other possibilities for the phase angles.  For example, if
$\phi_{x,0}+\phi_{x,f(x)}= f(x) \pi$ the scheme works with $S$
replaced by the identity operation.  Given arbitrary phase angles,
the scheme will work if the second oracle call is to the time
reversed oracle, $U^{-1}$.  However, if $U^{-1}$ is not available
then random phase angles defeat this scheme for quantum parallelism. 

A quantum computer can also make explicit use of the phase
information provided by an oracle.  For example, suppose we wish to
encode a function $h:\Sigma^n\rightarrow \{0,1\}$.  One way to do
this is to let $\phi_{x,0}=(h(x)+1) \pi$.  For simplicity suppose the
classical information is trivial, $f(x) \equiv 0$.  A classical
computer submitting a query to this oracle always receives the
uninteresting answer `$0$' but a quantum computer may easily read
$h(x)$.  One method for doing this rests on the added assumption that
there is a special string, $z$, $h(z)=1$.  The quantum computer is
prepared in the superposition, $|\psi\rangle$,  
\begin{equation}
|\psi\rangle=\frac{1}{\sqrt{2}}(|z,0\rangle +
|x,0\rangle)
\end{equation}
 This state is fed to the oracle which returns
\begin{equation}
|\chi\rangle = \frac{1}{\sqrt{2}}(|z,0\rangle +
(-1)^{h(x)+1}|x,0\rangle)
\end{equation}
 The inner product $\langle \psi | \chi
\rangle$ is  $0$ if 
$h(x)=0$ and $1$ if $h(x)=1$. Thus a measurement of $| \psi
\rangle\langle \psi |$ gives the value of $h(x)$.  

This type of oracle yields the ultimate relativized
separation between quantum and classical computation. Suppose $x$
enumerates Turing machines and let $h(x)=1$  if Turing machine $x$
halts and zero otherwise. A quantum computer plus the oracle
solves the halting problem whereas a classical computer is not
helped at all by the oracle. Thus useful relativized separations
between quantum and classical complexity classes must exclude the
transmission of phase encoded information from
oracle to computer. 

In conclusion, the complete definition of a quantum oracle must
include a specification of phase.   Phase encoded information can
be transmitted from an oracle to a quantum computer but not to 
classical computer using the same oracle device.   This raises the
interesting prospect that a quantum computational device
interacting with an environment may be more powerful than any
classical device in the same environment.

\end{document}